\documentclass[aps,prc,twocolumn,showpacs,preprintnumbers,nofootinbib,float,superscriptaddress,longbibliography]{revtex4-2}
\usepackage{graphicx, fancybox}
\usepackage{amsmath,amssymb}
\usepackage[colorlinks=true, pdfstartview=FitV, linkcolor=red, citecolor=blue, urlcolor=blue]{hyperref}
\usepackage{xcolor}
\usepackage{url}

\newcommand{\Rshear}{\tilde{R}_\pi}
\newcommand{\Rbulk}{\tilde{R}_\Pi}
\newcommand{\snn}{\sqrt{s_\mathrm{NN}}}

\usepackage[normalem]{ulem}

\begin{document}
\title{A Resummed Hydrodynamic Description of Relativistic Heavy-ion Collisions}

\author{Cheng Chiu}
\email{ccchiu@umich.edu}
\affiliation{Physics Department, University of Michigan, Ann Arbor, MI 48109, USA}

\author{Gabriel Denicol}
\email{gsdenicol@id.uff.br}
\affiliation{Instituto de F\'{ı}sica, Universidade Federal Fluminense Av. Gal. Milton Tavares de Souza, S/N, 24210-346, Gragoat\'{a}, Niter\'{o}i, Rio de Janeiro, Brazil}

\author{Matthew Luzum}
\email{mluzum@usp.br}
\affiliation{Instituto de F\'{i}sica, Universidade de S\~{a}o Paulo, R. do Mat\~{a}o, 1371, S\~{a}o Paulo, Brazil, 05508-090}

\author{Chun Shen}
\thanks{corresponding author}
\email{chunshen@wayne.edu}
\affiliation{Department of Physics and Astronomy, Wayne State University, Detroit, Michigan, 48201, USA}

\begin{abstract}
We introduce a resummed hydrodynamic scheme for evolving the viscous stress tensors in relativistic viscous hydrodynamics, based on which the necessary non-linear causality conditions can be imposed. When the magnitudes of the shear and bulk viscous stress tensors are small relative to the ideal part of the energy-momentum tensor, this new resummed scheme reduces to the standard second-order relativistic hydrodynamic theories. Nontrivial nonlinear corrections from high-order gradient terms retain the sizes of shear and bulk viscous stress tensors within tunable maximum allowed values. We perform event-by-event simulations for Pb+Pb and p+Pb collisions at 5.02 TeV to quantify the theoretical uncertainties from this resummed scheme on final-state flow observables.
\end{abstract}

{\maketitle}

\section{Introduction}

Relativistic viscous hydrodynamics is a long-wavelength effective framework to describe the macroscopic dynamics of the hot and dense phase of heavy-ion collisions created at the Relativistic Heavy Ion Collider (RHIC) and the Large Hadron Collider (LHC) \cite{Heinz:2013th, Gale:2013da, Shen:2020gef, Shen:2020mgh, Elfner:2022iae, Heinz:2024jwu}. The strongly-coupled quark-gluon plasma (QGP) created in heavy-ion collisions behaves like a liquid with a very small viscosity-to-entropy ratio~\cite{Song:2010mg, Shen:2015msa, Bernhard:2019bmu, Shen:2023awv}.
Event-by-event hydrodynamic simulations can accurately describe the time evolution of the QGP with well-defined equations of state from lattice QCD~\cite{Monnai:2021kgu, Monnai:2024pvy, Abuali:2025tbd} and calibrated specific shear and bulk viscosities~\cite{Bernhard:2019bmu, JETSCAPE:2020mzn, Nijs:2020roc, Jahan:2024wpj, Jahan:2025cbp}. These simulations can provide valuable insights into the QGP's transport coefficients, such as its viscosity and thermal conductivity, as well as the properties of the medium, including its temperature and energy density~\cite{Muller:2025qof}.

Although the QGP shares the common pressure-driven dynamics with other types of fluids in nature, it is unique due to its relativistic dynamics. The flow velocity of QGP in heavy-ion collisions can reach significant fractions of the speed of light when special relativistic effects are essential during its hydrodynamic expansion. Therefore, causality conditions are particularly important in modeling heavy-ion collisions due to the system's highly dynamic and rapidly evolving nature. Large pressure gradients can arise during the early stages of collisions, leading to potential causality violations if not properly accounted for. When dissipation is present, causality is also directly related to the system's stability~\cite{Gavassino:2021owo}.

Recently, flow-like signatures have been observed in small collision systems, such as proton+nucleus and proton+proton collisions~\cite{Nagle:2018nvi, Schenke:2021mxx, Noronha:2024dtq}. Due to their small system sizes, strong pressure gradients can drive the system far out of thermal equilibrium, generating significant dissipative effects. The quantitative understanding of collective behavior in these small collision systems has been a central topic in understanding how the QGP's strongly coupled nature emerged as a function of collision system size. Therefore, a robust formulation of causal relativistic viscous fluid dynamics is critical for theoretically interpreting the small system collisions at RHIC and the LHC~\cite{Shen:2020mgh, Noronha:2024dtq, Zhao:2022ayk, Zhao:2022ugy, Ryu:2023bmx, Schenke:2024cnc}. Quantifying theoretical uncertainties from causality conditions also becomes a critical issue when performing precision comparisons with experimental flow measurements~\cite{Jia:2022ozr, Giacalone:2024ixe, Domingues:2024pom, Mantysaari:2024uwn, Mantysaari:2025tcg, Zhang:2025hvi, Horecny:2025mak, Giacalone:2025vxa}. 

Formulating causal relativistic viscous fluid dynamics has been a long-standing problem~\cite{Hiscock:1983zz, Olson:1989ey,Pu:2009fj,Huang:2010sa,Denicol:2008ha}. When the collision systems evolve out of equilibrium, local dissipative contributions to the energy-momentum tensor can become comparable to the equilibrium pressure $P$. In these regions, the constraints on the transport coefficients derived in \cite{Hiscock:1983zz,Olson:1989ey,Pu:2009fj} using linearized perturbations around equilibrium are insufficient to ensure a well-defined causal evolution. The full non-linear causality conditions \cite{Bemfica:2020xym} have recently been derived for second-order Denicol-Niemi-Moln\'{a}r-Rischke (DNMR) hydrodynamics theory~\cite{Denicol:2012cn}. These non-linear causality conditions can be fulfilled by imposing a global constraint on the maximum-allowed inverse Reynolds numbers \cite{Chiu:2021muk, Plumberg:2021bme}. However, the equations of motion for conventional second-order hydrodynamic theories allow the size of the viscous tensor to increase indefinitely, leading to uncontrollable violation of causality conditions when large velocity gradients are present.

In this work, we develop a new phenomenological resummed hydrodynamic scheme to evolve the system's viscous stress tensors. It promotes the transport coefficients as functions of the local viscous stress tensors, which can be interpreted as resumming an infinite set of gradient terms in the gradient expansion. This numerically agnostic scheme introduces essential nonlinearity to impose upper bounds on the system's inverse Reynolds numbers. Similar works were proposed for the bulk viscosity sector only~\cite{Gavassino:2023xkt, Yang:2023ogo}.
Based on this feature, we can enforce the necessary causality conditions on individual fluid cells. The resummed scheme automatically ensures that these conditions are satisfied during the evolution once the constraints are imposed on the initial conditions.

As the first phenomenological application, we perform event-by-event hydrodynamic simulations for Pb+Pb and p+Pb collisions at $\sqrt{s_\mathrm{NN}} = 5.02$~TeV and quantify the effects of the resummed numerical scheme on final-state flow observables as an estimate of theoretical uncertainty. 

\section{A Resummed Hydrodynamic Framework}
\label{sec:resummedScheme}

The second-order relativistic viscous hydrodynamics equations (without any conserved charges) are,
\begin{align}
        \partial_\mu T^{\mu\nu} &= 0, \label{eq:ideal} \\
	\tau_{\Pi} \dot{\Pi} + \Pi
	& = - \zeta \theta
	- \delta_{\Pi \Pi} \Pi \, \theta
	+ \lambda_{\Pi \pi} \pi^{\mu\nu} \sigma_{\mu\nu} \label{eq:bulk} \\ 
	\tau_{\pi} \dot{\pi}^{\langle \mu\nu \rangle}
	+ \pi^{\mu\nu}
	& = 2\eta \sigma^{\mu\nu}
	- \delta_{\pi \pi} \pi^{\mu\nu} \theta
	+ \varphi_7 \pi_{\alpha}^{\langle \mu} \pi^{\nu\rangle \alpha} \nonumber \\
	& \quad - \tau_{\pi \pi} \pi_{\alpha}^{\langle \mu} \sigma^{\nu\rangle \alpha}
	+ \lambda_{\pi \Pi} \Pi\, \sigma^{\mu\nu}\,.
	\label{eq:shear}
\end{align}
The shear ($\eta$) and bulk ($\zeta$) viscosities are free transport parameters, and Table~\ref{table:transport_coeffs} lists the standard choice of the other transport coefficients for the DNMR hydrodynamic theory in Eqs.~\eqref{eq:bulk} and \eqref{eq:shear}.

\begin{table}[h!]
\caption{The choice of second-order transport coefficients used in the full DNMR hydrodynamic theory \cite{Denicol:2014vaa}.}
\begin{tabular}{c|c|c|c|c|c}
\hline \hline
     & $\frac{\zeta}{\tau_\Pi (\varepsilon + P)}=\frac{1}{C_\zeta}$ & $\frac{\delta_{\Pi\Pi}}{\tau_\Pi}$ & $\frac{\lambda_{\Pi\pi}}{\tau_\Pi}$ & & \\ \hline
     DNMR (bulk) & $14.55(\frac{1}{3}-c_s^2)^2$ &  $\frac{2}{3}$ & $\frac{8}{5}(\frac{1}{3} - c_s^2)$ & & \\ \hline
     & $\frac{\eta}{\tau_\pi (\varepsilon + P)}=\frac{1}{C_\eta}$ & $\frac{\delta_{\pi\pi}}{\tau_\pi}$  & $\frac{\tau_{\pi\pi}}{\tau_\pi}$ & $\frac{\lambda_{\pi\Pi}}{\tau_\pi}$ & $\varphi_7$ \\ \hline
     DNMR (shear) & $\frac{1}{5}$ & $\frac{4}{3}$ &  $\frac{10}{7}$ & $\frac{6}{5}$ &  $\frac{9}{70} \frac{4}{\varepsilon + P}$ \\ \hline \hline
\end{tabular}
\label{table:transport_coeffs}
\end{table}

In hydrodynamic simulations, it is practical to track the evolution of the inverse Reynolds numbers for the shear stress tensor and bulk viscous pressure to assess how far the system is away from thermal equilibrium.
We define the following inverse Reynolds numbers related to shear and bulk viscous stress tensors,
\begin{align}
    \Rshear &\equiv \frac{2}{\sqrt{6}}\frac{\sqrt{\pi^{\mu\nu} \pi_{\mu\nu}}}{\varepsilon + P} \label{eq:Rshear} \\
    \Rbulk &\equiv \frac{\vert \Pi \vert}{\varepsilon + P}. \label{eq:Rbulk}
\end{align}
The definition of $\Rshear$ in Eq.~\eqref{eq:Rshear} ensures $\Rshear \ge \mathrm{max}\{ |\Lambda_i| \}$, where $\Lambda_i (i = 0 - 3)$ are the eigenvalues of the shear viscous tensor~\cite{Chiu:2021muk}.

Using the shear and bulk inverse Reynolds numbers, we can simplify the full necessary causality conditions derived in Ref.~\cite{Bemfica:2020xym} into the following two inequalities:
\begin{align}
    1 - \Rbulk + \left(\frac{\tau_{\pi\pi}}{\tau_\pi} - 1 \right) \Rshear \ &\geq N_1 \geq 0
    \label{eq:N1_range} \\
    1 - \Rbulk + \Rshear  &\geq N_2 \geq 0,  \label{eq:N2_range}
\end{align}
with
\begin{align}
    N_1 &\equiv \frac{1}{C_\eta} - \frac{\lambda_{\pi\Pi}}{2\tau_\pi} \Rbulk - \frac{\tau_{\pi\pi}}{4\tau_\pi}\Rshear
    \label{eq:N1} \\
    N_2 &\equiv c^2_s + \frac{4}{3}\frac{1}{C_\eta} + \frac{1}{C_\zeta} - \left(\frac{2}{3}\frac{\lambda_{\pi\Pi}}{\tau_\pi} + \frac{\delta_{\Pi\Pi}}{\tau_\Pi} + c^2_s \right) \Rbulk \nonumber \\
    & \quad + \left( \frac{3\delta_{\pi\pi} + \tau_{\pi\pi}}{3\tau_\pi} + \frac{\lambda_{\Pi\pi}}{\tau_\Pi} + c^2_s\right) \Rshear.
    \label{eq:N2}
\end{align}

Please note that because we reduce the variable dimensions of the causality conditions, the simplified necessary causality conditions, Eqs.~\eqref{eq:N1_range} and \eqref{eq:N2_range}, impose more stringent constraints on the values of shear stress tensor and bulk viscous pressure than those from the original necessary causality conditions in Ref.~\cite{Bemfica:2020xym}.

A system with large values of inverse Reynolds numbers will violate the necessary causality conditions. In our resummed hydrodynamic scheme, we want to introduce modifications to the transport coefficients in Eq.~\eqref{eq:bulk} and \eqref{eq:shear} when the system is far from thermal equilibrium. By promoting the transport coefficients as functions of the local inverse Reynolds numbers, we can introduce dynamical corrections to the original PDE such that the shear and bulk inverse Reynolds numbers are bounded during the hydrodynamic evolution. Using this feature, we can ensure the system fulfills the necessary causality conditions by constraining the absolute magnitudes of $\Rshear$ and $\Rbulk$.

\begin{figure*}[ht!]
  \centering
  \begin{tabular}{cc}
    \includegraphics[width=0.48\linewidth]{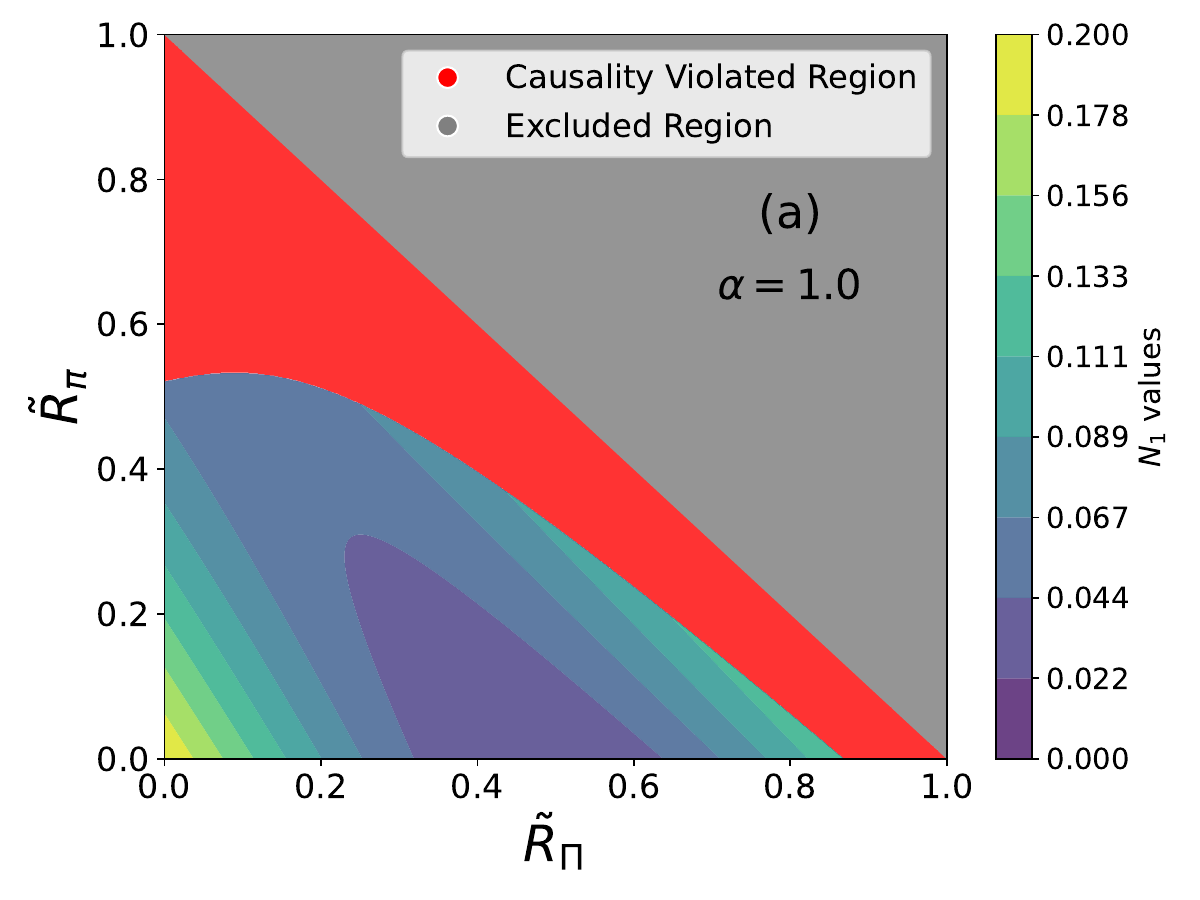} &
    \includegraphics[width=0.48\linewidth]{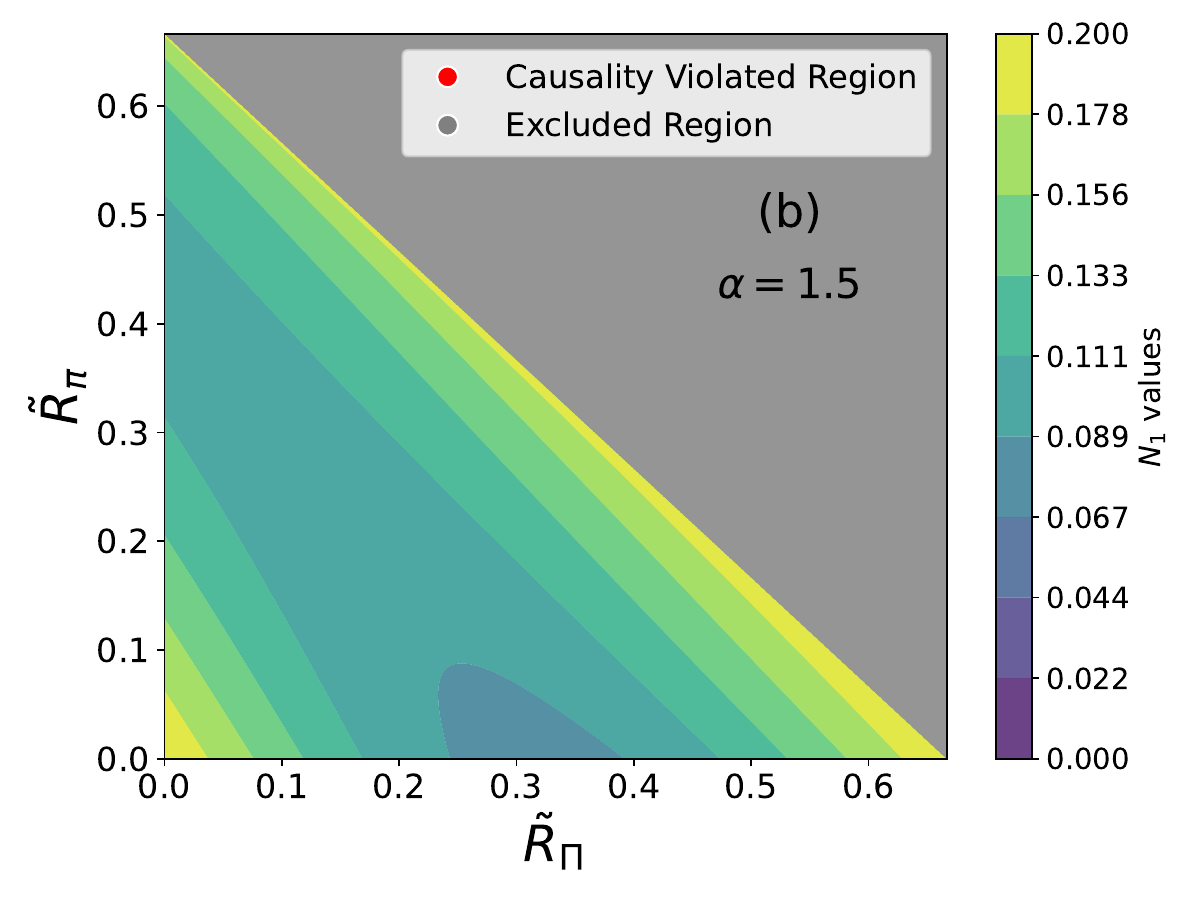} \\
    \includegraphics[width=0.48\linewidth]{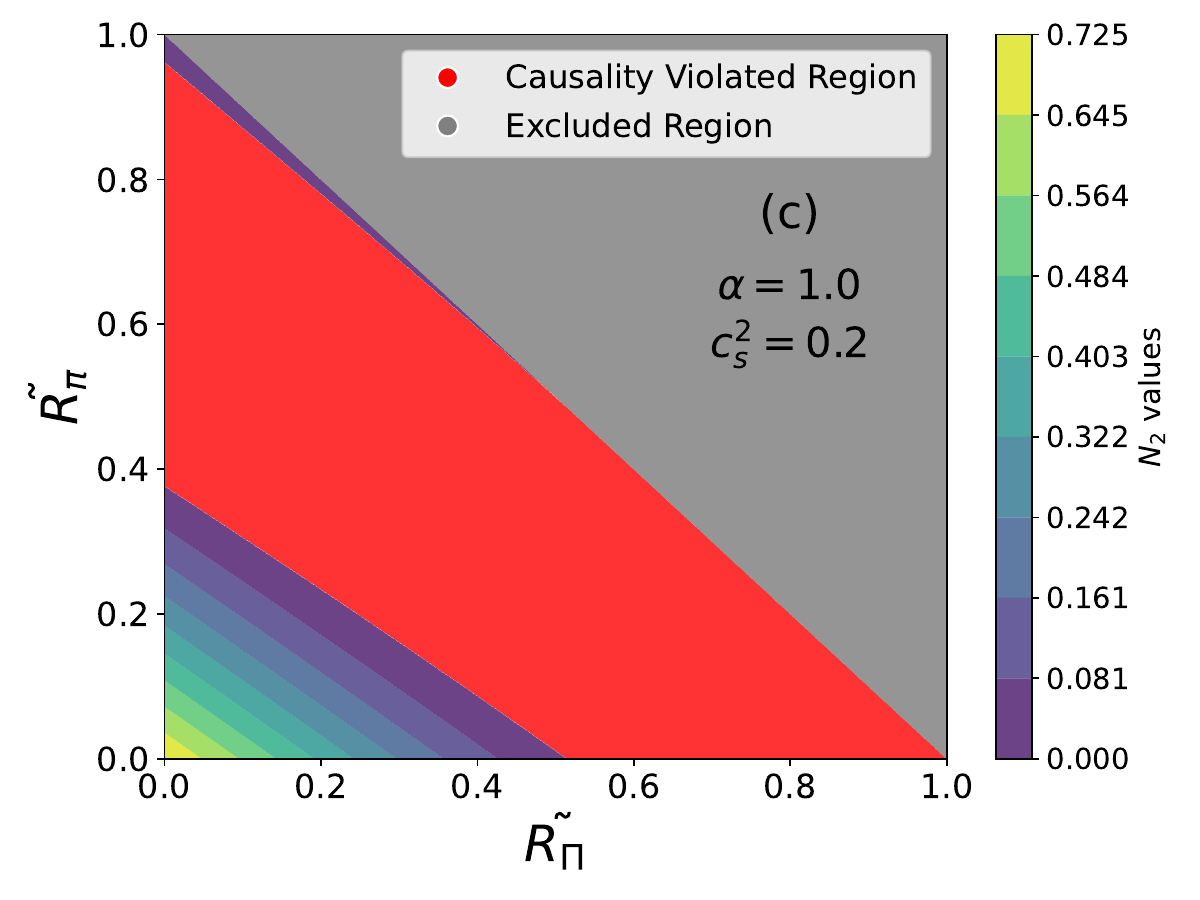} &
    \includegraphics[width=0.48\linewidth]{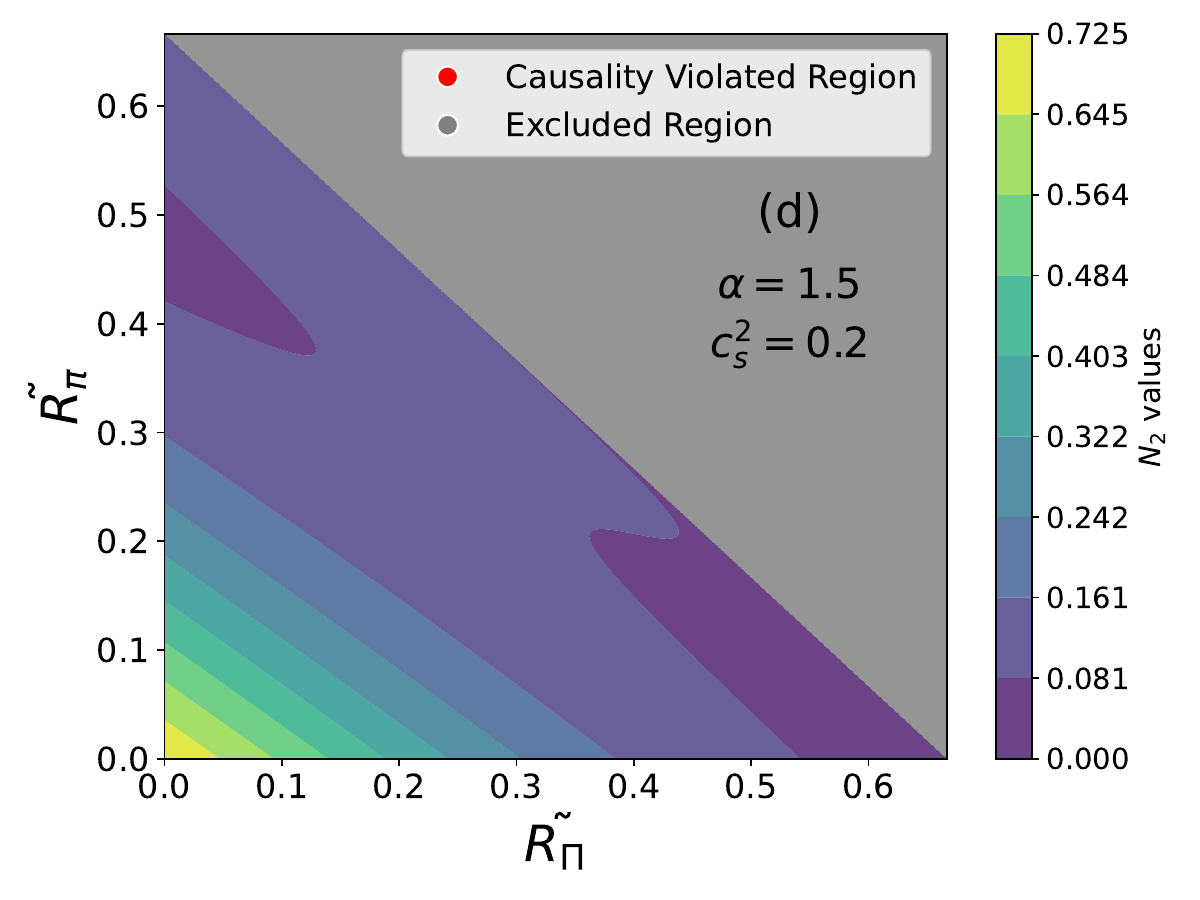}
  \end{tabular}
  \caption{Color contour plots visualize Eq.~\eqref{eq:N1_range} (panel a and b) and \eqref{eq:N2_range} (panel c and d) in the $\Rbulk$ and $\Rshear$ parameter space. It includes three regions: satisfied, violated (red), and excluded region (gray upper triangle). We present the necessary causality conditions with a typical value of speed of sound square $c_s^2 = 0.2$.}
  \label{fig:N}
\end{figure*}

We propose the following renormalization scheme to scale the transport coefficients on the right-hand sides of Eqs.~\eqref{eq:bulk} and \eqref{eq:shear}:
\begin{align}
    & \eta \rightarrow f \eta,\quad \frac{\delta_{\pi\pi}}{\tau_\pi} \rightarrow f^2\frac{\delta_{\pi\pi}}{\tau_\pi},\quad \frac{\tau_{\pi\pi}}{\tau_\pi} \rightarrow f^2\frac{\tau_{\pi\pi}}{\tau_\pi}, \\
    & \hspace{1.6cm} \frac{\lambda_{\pi \Pi}}{\tau_{\pi}} \rightarrow f^2\frac{\lambda_{\pi \Pi}}{\tau_{\pi}},\quad \frac{\varphi_7}{\tau_{\pi}} \rightarrow f \frac{\varphi_7}{\tau_{\pi}} \\
    & \zeta \rightarrow f \zeta,\quad \frac{\delta_{\Pi\Pi}}{\tau_\Pi} \rightarrow f\frac{\delta_{\Pi\Pi}}{\tau_\Pi},\quad \frac{\lambda_{\Pi\pi}}{\tau_\Pi} \rightarrow f\frac{\lambda_{\Pi\pi}}{\tau_\Pi}
\end{align}
where 
\begin{equation}
    f \equiv \frac{1}{1 + \mathrm{arctanh}^2(\alpha (\Rbulk + \Rshear))}.
    \label{eq:renormf}
\end{equation}
We keep the relaxation times $\tau_\pi$ and $\tau_\Pi$ unchanged.
In Eq.~\eqref{eq:renormf}, the parameter $\alpha > 0$ controls the maximum allowed value for $\Rbulk + \Rshear$ because the $\mathrm{arctanh}$ function is only properly defined when its argument $\alpha (\Rbulk + \Rshear) \in [-1, 1]$. The choice of $\mathrm{arctanh}$ functional dependence is not motivated from any microscopic theory, but for numerical convenience at the current stage. 
When $\Rbulk + \Rshear \rightarrow 1/\alpha$, $f \rightarrow 0$. This renormalization factor suppresses all the transport coefficients on the right-hand sides of Eqs.~\eqref{eq:bulk} and \eqref{eq:shear} to zero. In this case, the equations for shear and bulk viscous tensor become,
\begin{align}
    \dot{\Pi} &= -\frac{1}{\tau_\Pi} \Pi \\
    \dot{\pi}^{\langle \mu \nu \rangle} &= -\frac{1}{\tau_\pi} \pi^{\mu\nu}.
\end{align}
In this limit, the magnitude of $\Pi$ and $\pi^{\mu\nu}$ will decay exponentially. We have checked that the evolution of $\Rshear$ and $\Rbulk$ is safely bounded by the allowed maximum value, namely $\Rbulk + \Rshear \le 1/\alpha$, in practice.

\begin{figure}[t!]
  \centering
    \includegraphics[width=0.9\linewidth]{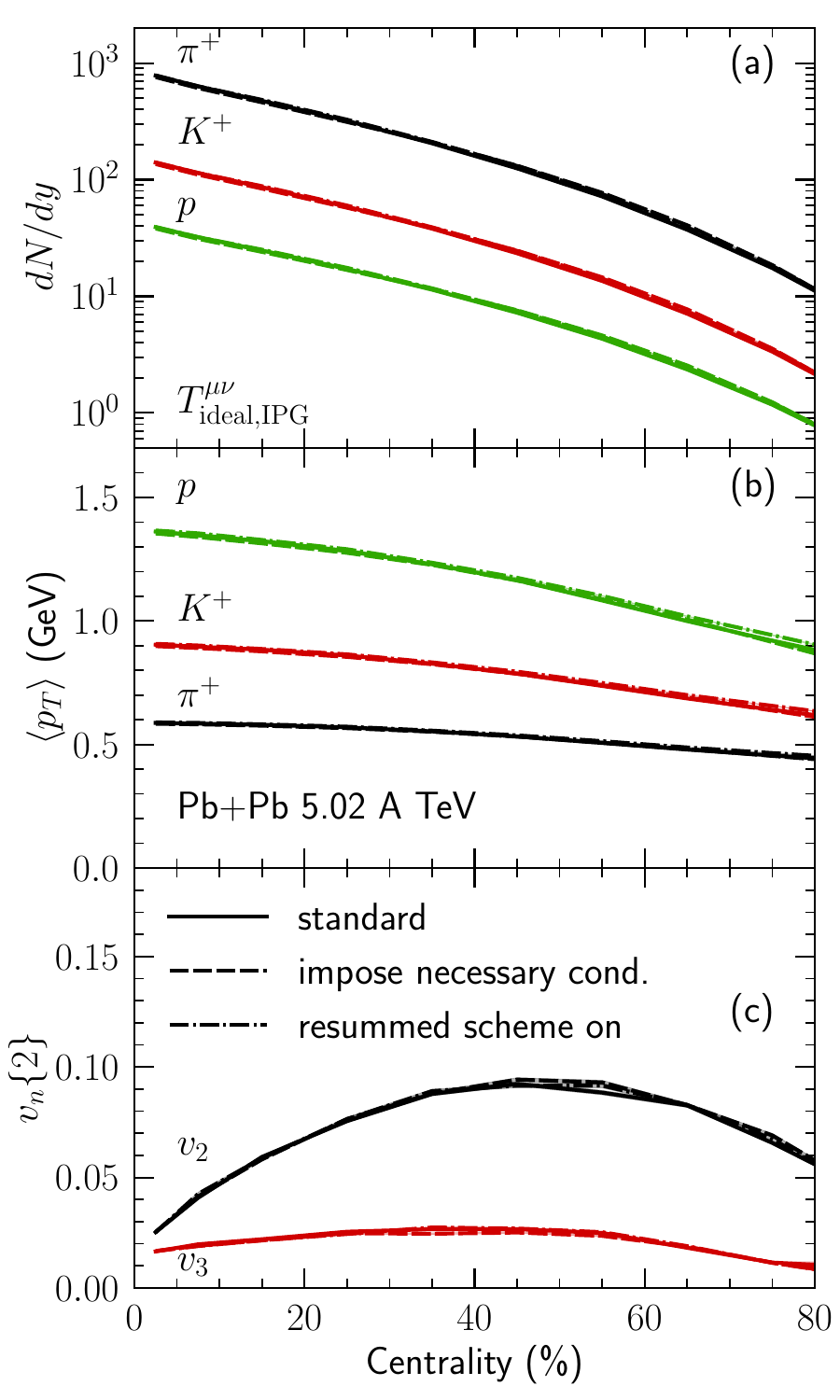}
  \caption{Simulations of identified particle yields, their averaged transverse momenta, and charged hadron anisotropic flow coefficients in Pb+Pb collisions at $\snn = 5.02$ TeV. The figure compares different regulation methods when hydrodynamic fields are initialized with the ideal part of the energy-momentum tensor from the IP-Glasma model. The colored bands indicate statistical errors in the simulation results.}
  \label{fig:PbPbzero}
\end{figure}
\begin{figure}[t!]
  \centering
    \includegraphics[width=0.9\linewidth]{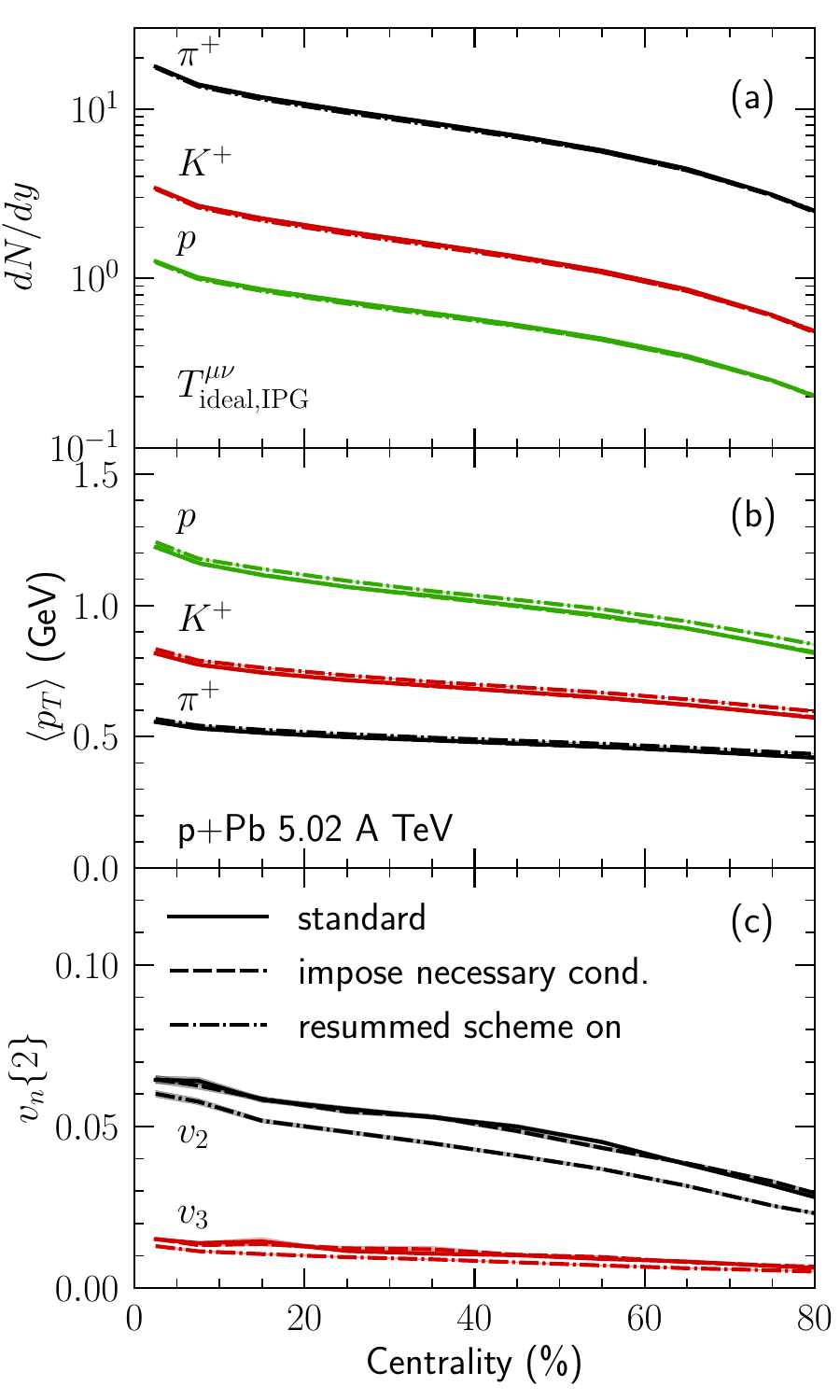}
  \caption{Similar to Fig.~\ref{fig:PbPbzero} but for mid-rapidity observables in p+Pb collisions at $\snn = 5.02$~TeV as functions of collision centrality.}
  \label{fig:pPbzero}
\end{figure}

For small inverse Reynolds numbers, we can expand the renormalization factor as,
\begin{align}
    f \approx 1 - \alpha^2 (\Rbulk + \Rshear)^2.
\end{align}
To consider a simplified case where only bulk viscous pressure is non-zero,
\begin{align}
    \dot{\Pi} &= -\frac{1}{\tau_\Pi} (\Pi + f \zeta \theta) - f \frac{\delta_{\Pi \Pi}}{\tau_\Pi} \Pi \theta \nonumber \\
    &\approx  -\frac{1}{\tau_\Pi} (\Pi + \zeta \theta) - \frac{\delta_{\Pi \Pi}}{\tau_\Pi} \Pi \theta \nonumber \\
    & \quad + \frac{\zeta}{\tau_\Pi} \alpha^2 \Rbulk^2 \theta + 2 \frac{\delta_{\Pi \Pi}}{\tau_\Pi} \alpha^2 \Rbulk^2 \Pi \theta.
\end{align}
Considering $\Pi \sim \zeta \theta$, the terms with first and second-order gradients are unmodified in the small inverse Reynolds number limit. The corrections to the equation of motion are entered in the third order in gradients. The Taylor expansion of the scaling factor $f$ in terms of $\Rshear$ and $\Rbulk$ determines the exact high-order gradient terms being resummed to correct the transport coefficients when the system is far from equilibrium. With the parameter $\alpha$ in the scaling factor $f$, we can effectively control the maximum allowed values of the system's inverse Reynolds numbers. In practice, we choose the value of $\alpha$ such that the system satisfies the non-linear necessary causality conditions.

We note that the qualitative behavior of the decreasing renormalization factor $f$ in the presence of large gradients or far-out-of-equilibrium agrees with some theoretical expectations as derived in Refs.~\cite{Gavassino:2023xkt, Romatschke:2017vte}. Meanwhile, the presence of thermal fluctuations may set a lower bound for this renormalization factor~\cite{Kovtun:2011np, Akamatsu:2016llw, Chattopadhyay:2025uqo}, which we will devote to future work.

Figure~\ref{fig:N} shows the necessary causality conditions in Eqs.~\eqref{eq:N1_range} and \eqref{eq:N2_range} as functions of the inverse Reynolds numbers $\Rshear$ and $\Rbulk$. We observed that with the choice of $\alpha = 1$ for the renormalization factor $f$ in Eq.~\eqref{eq:renormf}, there are allowed phase spaces for the shear and bulk inverse Reynolds numbers to violate the necessary causality conditions (indicated as red). With $\alpha = 1.5$, we verify that the necessary causality conditions are valid for all allowed values of $\Rshear$ and $\Rbulk$ for the speed of sound squared in the range $0.13 < c_s^2 < 1/3$, which is relevant for the lattice QCD equation of state. It constrained the maximum allowed values for $\Rshear$ and $\Rbulk$ to be 2/3.

\section{Phenomenological Effects of the Resummed Hydrodynamics on Observables}
\label{sec:results}

In this section, we will investigate the effects of the proposed resummed hydrodynamic scheme in simulating large and small collision systems at the $\sqrt{s_\mathrm{NN}} = 5.02$\,TeV.
We will compare simulation results from the following three setups. (i) The results from the ``standard'' approach only impose regulations on viscous stress tensors for numerical stability as described in Ref.~\cite{Shen:2014vra, Schenke:2020mbo}. (ii) For the ``impose necessary cond.'' case, we reduce the sizes of viscous stress tensors in every fluid cell when they violate the necessary causality condition (see the detailed procedure in the Appendix). (iii) The ``resummed scheme on'' case enables the resummed transport coefficients as described in Sec.~\ref{sec:resummedScheme}.

To explore the maximum phenomenological effects from the different numerical schemes, we perform tests using the IP-Glasma + MUSIC + UrQMD framework~\cite{Schenke:2020mbo, Schenke:2020unx, Paquet:2015lta}, in which the IP-Glasma initial conditions provides large pressure gradients and its energy-momentum tensor is far out of equilibrium when matching to hydrodynamics at $\tau_\mathrm{hydro} = 0.4$\,fm/$c$. The average inverse Reynolds number for the shear stress tensor is around $\Rshear \sim 0.6$~\cite{Schenke:2019pmk}. 

We will compare different numerical schemes on final-state observables by initializing hydrodynamic simulations at three levels of out-of-equilibrium, demonstrating sensitivity to the pre-equilibrium stage of evolution. 

\begin{figure}[t!]
  \centering
    \includegraphics[width=0.9\linewidth]{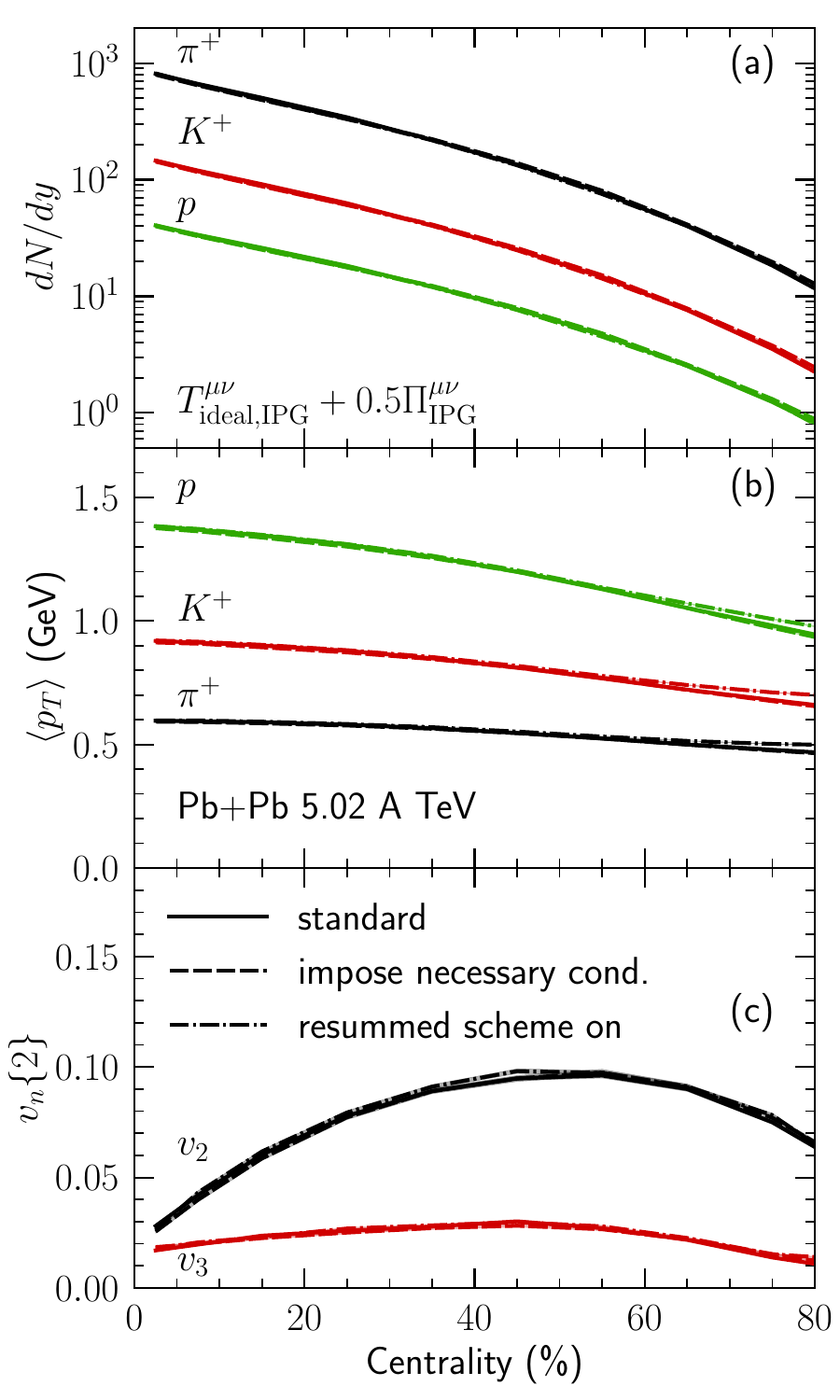}
  \caption{Similar to Fig.~\ref{fig:PbPbzero} but we initialize the hydrodynamics with $T^{\mu\nu}_\mathrm{hydro} = T^{\mu\nu}_\mathrm{ideal, IPG} + 0.5 \Pi^{\mu\nu}_\mathrm{IPG}$ from the IP-Glasma model.}
  \label{fig:PbPbhalf}
\end{figure}

\begin{figure}[t!]
  \centering
    \includegraphics[width=0.9\linewidth]{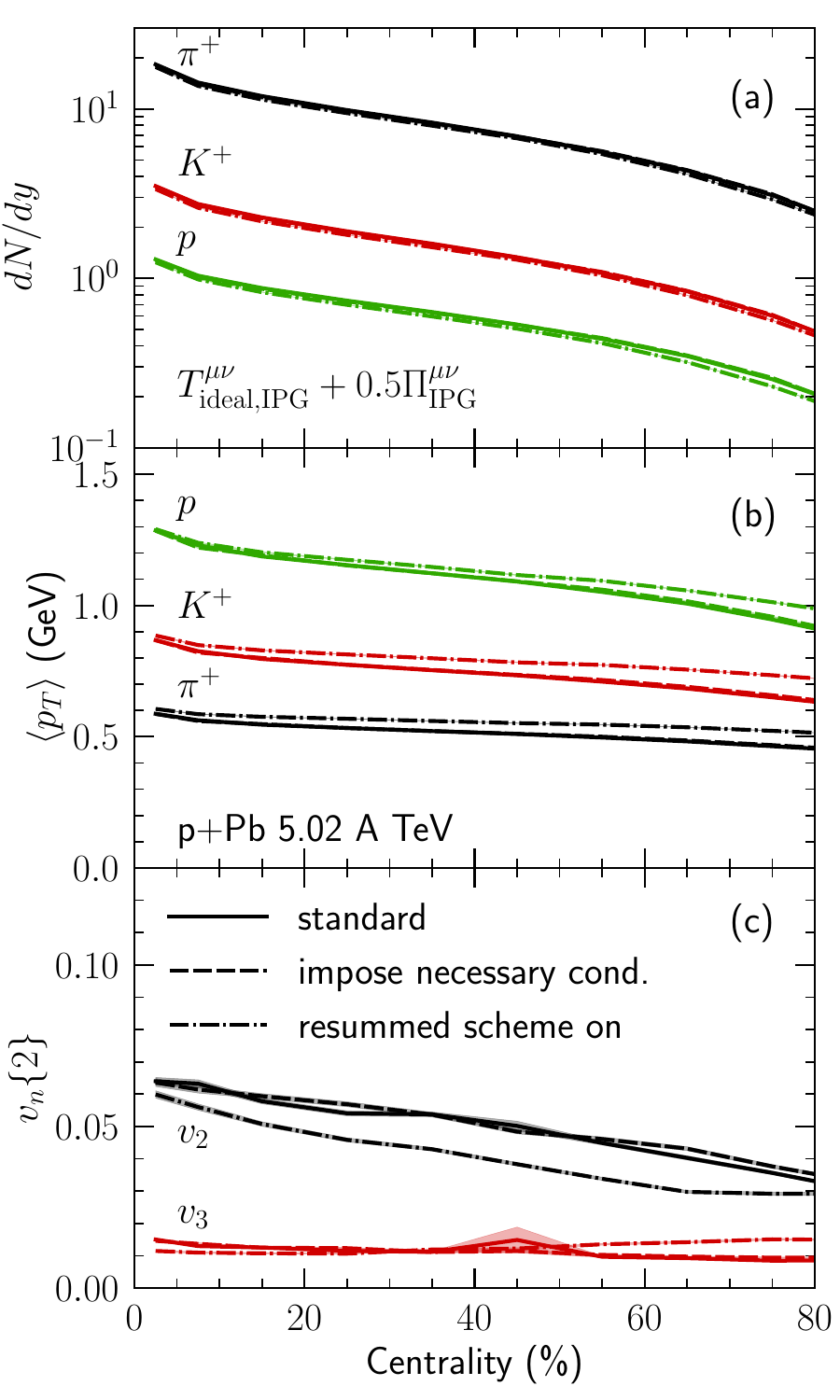}
  \caption{Similar to Fig.~\ref{fig:PbPbhalf} but for mid-rapidity observables in p+Pb collisions at $\snn = 5.02$~TeV.}
  \label{fig:pPbhalf}
\end{figure}

\subsection{Initializing hydrodynamics at equilibrium}
\label{sec:ResultsEq}

We start hydrodynamic simulations assuming the collision systems are in complete thermal equilibrium, namely $\Rshear = \Rbulk = 0$ for all fluid cells at $\tau_\mathrm{hydro}$.

Figures~\ref{fig:PbPbzero} and \ref{fig:pPbzero} show the identified particle yields, their averaged transverse momenta, and charged hadron anisotropic flow coefficients as functions of collision centrality for Pb+Pb and p+Pb collisions at 5.02 TeV.
In this case, the system needs some time to evolve out of equilibrium. Therefore, the violation of causality is minimal at early stages. The different numerical schemes give almost identical results for all the final-state observables in Pb+Pb collisions, indicating our proposed numerical regulations are well-behaved when the system starts from equilibrium.
In the meantime, we observe a 10\% difference in charged hadron anisotropic flow coefficients in the small-size p+Pb collisions.

\begin{figure}[t!]
  \centering
    \includegraphics[width=0.9\linewidth]{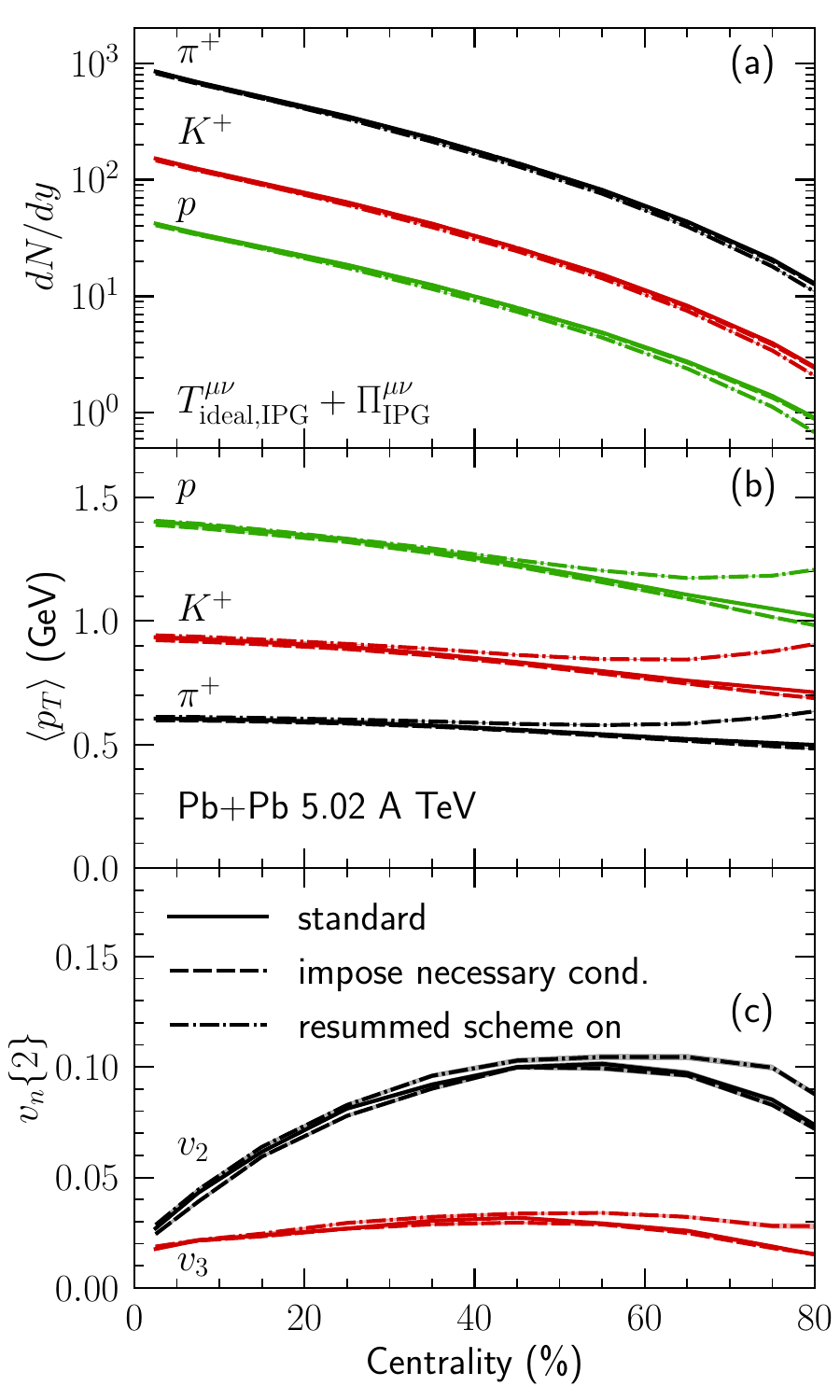}
  \caption{Similar to Fig.~\ref{fig:PbPbzero}, but we initialize hydrodynamics with the full energy-momentum tensor from the IP-Glasma model.}
  \label{fig:PbPbone}
\end{figure}

\begin{figure}[t!]
  \centering
    \includegraphics[width=0.9\linewidth]{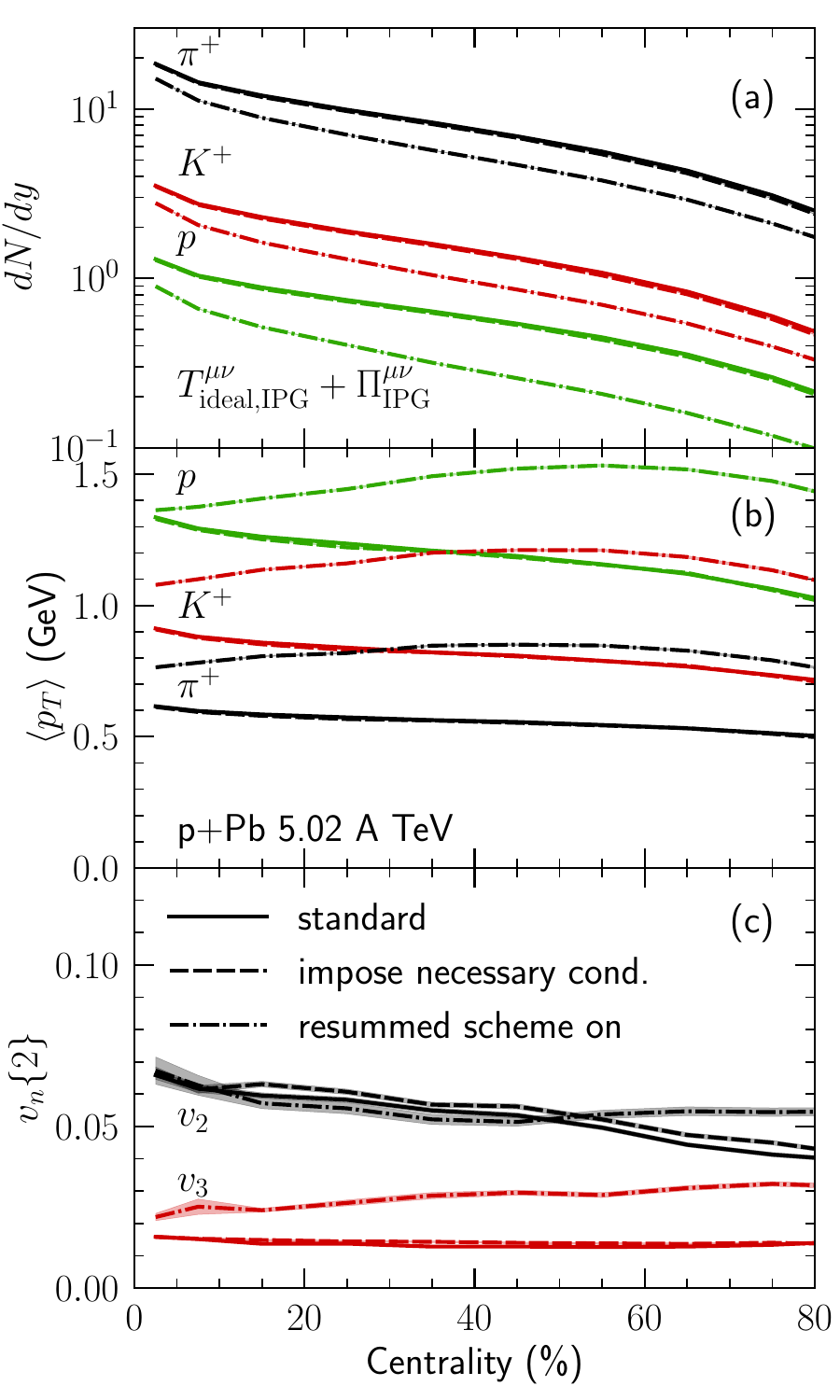}
  \caption{Similar to Fig.~\ref{fig:PbPbone} but for mid-rapidity observables in p+Pb collisions at 5.02 $A$ TeV.}
  \label{fig:pPbone}
\end{figure}

\subsection{Initializing hydrodynamics near equilibrium}
\label{sec:ResultsNearEq}

Now, we initialize our hydrodynamic simulations with half the size of the viscous stress tensor from the IP-Glasma model at $\tau_\mathrm{hydro} = 0.4$~fm/$c$.  In this case, the initial collision system is out-of-equilibrium, but the values $\Rshear + \Rbulk$ are within the allowed ranges for the proposed resummed hydrodynamic scheme.

Figures~\ref{fig:PbPbhalf} and \ref{fig:pPbhalf} compare the same final-state observables described in the previous section. Again, we observed negligible differences among the numerical schemes in Pb+Pb collisions. The results in Figs.~\ref{fig:PbPbzero} and \ref{fig:PbPbhalf} demonstrate that the proposed numerical regulation schemes only introduce negligible perturbative corrections to the original DNMR hydrodynamic theory when most of the simulated system evolves within the well-behaved phase space. For small collision systems, Fig.~\ref{fig:pPbhalf} shows a slightly larger numerical difference in the observables than in Fig.~\ref{fig:pPbzero}. This result indicates that the expansion in small systems is significant, as the system evolves out of equilibrium and triggers sizable numerical regulations.

\subsection{Initializing hydrodynamics far from equilibrium}
\label{sec:ResultsNoEq}

Lastly, we investigate the phenomenological impacts of the different numerical schemes when initializing hydrodynamic simulations with the full energy-momentum tensor from the IP-Glasma model. At the starting time $\tau_\mathrm{hydro}$, the values of $\Rshear$ and $\Rbulk$ in a fraction of cells are larger than the maximum allowed values for the proposed resummed scheme. In these cases, we regulate the values of the viscous stress tensors such that $\alpha(\Rshear + \Rbulk) = 1$ in all fluid cells. In these cases, the renormalization factor $f = 0$ in Eq.~\eqref{eq:renormf} and the system evolves as ideal hydrodynamics. An alternative approach for large inverse Reynolds numbers is to evolve the system using an effective kinetic theory and match to hydrodynamics later~\cite{Kurkela:2018wud, Gale:2021emg}.

Figures~\ref{fig:PbPbone} and \ref{fig:pPbone} show the comparisons of the three numerical schemes on final-state observables for large and small collision systems. Notably, even though there is a significant fraction of fluid cells that violate causality during the early time of the evolution in this setup~\cite{Chiu:2021muk, Plumberg:2021bme}, the results from the ``standard'' and ``impose necessary cond.'' are very close to each other for both Pb+Pb and p+Pb collisions. This comparison indicates that the numerical regulations applied to the causality-violating fluid cells do not have a substantial impact on the final-state observables.

Compared with the results from the resummed hydrodynamic scheme, we observe noticeable differences in peripheral collisions for centralities larger than 50\%. The final-state observables in central and semi-peripheral Pb+Pb collisions remain robust across different numerical schemes. This is because central Pb+Pb collisions have a long enough lifetime to wash out the impacts from resummed transport coefficients on the final-state observables. 

On the other hand, the peripheral Pb+Pb and small-size p+Pb collisions show strong sensitivity to the resummed hydrodynamic scheme when we initialize hydrodynamic simulations far from equilibrium. Because the inverse Reynolds numbers of the system are large, the proposed resummed hydrodynamic scheme will evolve systems with low effective viscosities, resulting in reduced viscous entropy production and large radial and anisotropic flows.
Therefore, the results using the full viscous tensor from the IP-Glasma model yield smaller particle yields and larger identified-particle mean $p_T$ and charged-hadron $v_n$ coefficients in Figs.~\ref{fig:PbPbone} and ~\ref{fig:pPbone} than those with half of the initial viscous tensor shown in Sec.~\ref{sec:ResultsNearEq}.

\section{Conclusion}
\label{sec:conclusion}

In this work, we propose a resummed hydrodynamic scheme, in which the effective transport coefficients depend on the values of the local shear stress tensor $\pi^{\mu\nu}$ and bulk viscous pressure $\Pi$ of the fluid cells. These renormalization corrections to the transport coefficients can be interpreted as a result of the resummation of a specific set of high-order gradient terms.
The goal of our work is not to derive a new hydrodynamic theory from a microscopic description, but rather to have an agnostic numerical framework to investigate the impacts of resumming large gradient terms on heavy-ion phenomenology. 
On the one hand, the proposed resummed equations of motion for the viscous stress tensors reduce to the regular DNMR hydrodynamic theory at the limit of thermal equilibrium. On the other hand,  when the system is far from equilibrium, the transport coefficients are renormalized to zero, quenching the viscous stress tensors exponentially. As a result, this resummed hydrodynamic scheme allows us to impose flexible numerical upper bounds for the inverse Reynolds' numbers defined in Eqs.~\eqref{eq:Rshear} and \eqref{eq:Rbulk}, namely $\Rshear + \Rbulk \le 1/\alpha$ ($\alpha$ is defined in Eq.~\eqref{eq:renormf}). Using such flexibility, we find that $\alpha = 1.5$ can effectively impose the non-linear necessary causality conditions for relativistic viscous hydrodynamics. 

We implement the proposed resummed hydrodynamic scheme in one of the state-of-the-art hydrodynamic frameworks, MUSIC. We quantify its impacts on the final-state observables in Pb+Pb and p+Pb collisions at the LHC energy using the IP-Glasma + MUSIC + UrQMD framework. We find that the resummed hydrodynamic scheme gives negligible corrections to the results from the standard DNMR theory when hydrodynamic simulations are initialized near thermal equilibrium. For hydrodynamic simulations of small systems initialized far from equilibrium, the resummed hydrodynamic scheme leads to close to ideal hydrodynamic evolution. The difference between the different numerical schemes indicates significant theoretical uncertainty in modeling the collective dynamics in small collision systems at high energies.

Because we can impose flexible upper bounds on the system's inverse Reynolds numbers with the parameter $\alpha$ in Eq.~\eqref{eq:renormf}, the proposed resummed hydrodynamic scheme can serve as an effective numerical framework for incorporating theoretical inputs from microscopic theories in the future. Such an agnostic numerical framework can help us derive data-driven constraints on the form of resummed hydrodynamics via Bayesian inference.

This work's simulation data, numerical code framework, and parameter files are available~\cite{chiu_2025_16780467}.

\section*{Acknowledgments}
We gratefully acknowledge the Galileo Galilei Institute for Theoretical Physics (GGI) in Florence for its hospitality and for hosting the workshop ``Foundations and Applications of Relativistic Hydrodynamics,'' which provided a stimulating environment and valuable discussions that contributed to this work.
This work is supported in part by the U.S. Department of Energy, Office of Science, Office of Nuclear Physics, under DOE Award No. DE-SC0021969, in part by grant NSF PHY-2309135 to the Kavli Institute for Theoretical Physics (KITP).
C.S. acknowledges a DOE Office of Science Early Career Award. 
This research was done using computational resources provided by the Open Science Grid (OSG)~\cite{Pordes:2007zzb, Sfiligoi:2009cct, OSPool, OSDF}, which is supported by the National Science Foundation awards \#2030508 and \#2323298.
M.L.~was supported by FAPESP projects 2017/05685-2, 2018/24720-6, and 2023/13749-1, by project INCT-FNA Proc.~No.~464898/2014-5, and by CAPES - Finance Code 001.

\appendix*
\section{An Alternative Numerical Regulation Method}
\label{sec:numericalReg}

While the proposed resummed framework can ensure the necessary causality conditions, it potentially can over-correct the original viscous stress tensor because we replace individual eigenvalues of the shear stress tensor by $\Rshear$ in Eqs.~\eqref{eq:N1_range} and \eqref{eq:N2_range}. To test its performance near the causality violation limits, we implement an alternative numerical scheme that regulates the viscous stress tensors at every step of hydrodynamic evolution to satisfy the original necessary causality conditions in Ref.~\cite{Bemfica:2020xym}.

According to Ref.~\cite{Bemfica:2020xym, Chiu:2021muk}, causality is violated when one applies the DNMR hydrodynamic equation of motion to a system which is far out of equilibrium; equivalently, it is when the magnitude of bulk pressure $\Pi$ and/or shear stress tensor $\sqrt{\pi^{\mu\nu} \pi_{\mu\nu}}$ is too large and violates at least one of the following causality measures $\{n_i\} (i = 1 - 6)$.
\begin{align}
    n_1 \equiv \frac{1}{C_\eta}+\frac{\lambda_{\pi\Pi}}{2\tau_\pi}\frac{\Pi}{\varepsilon+P}-\frac{\tau_{\pi\pi}}{4\tau_\pi}\frac{|\Lambda_1|}{\varepsilon+P} \geq 0,
    \label{eq:causal_n1}
\end{align}
\begin{align}
    n_2 \equiv 1 - \frac{1}{C_\eta}+\left(1-\frac{\lambda_{\pi\Pi}}{2\tau_\pi}\right)\frac{\Pi}{\varepsilon+P}-\frac{\tau_{\pi\pi}}{4\tau_\pi}\frac{\Lambda_3}{\varepsilon+P} \geq 0,
    \label{eq:causal_n2}
\end{align}
\begin{align}
    n_3 \equiv \frac{1}{C_\eta}+\frac{\lambda_{\pi\Pi}}{2\tau_\pi}\frac{\Pi}{\varepsilon+P}-\frac{\tau_{\pi\pi}}{4\tau_\pi}\frac{\Lambda_3}{\varepsilon+P} \geq 0,
    \label{eq:causal_n3}
\end{align}
\begin{align}
   n_4 &\equiv 1 - \frac{1}{C_\eta} + \left(1-\frac{\lambda_{\pi\Pi}}{2\tau_\pi}\right)\frac{\Pi}{\varepsilon+P} \nonumber \\
    & + \left(1-\frac{\tau_{\pi\pi}}{4\tau_\pi} \right) \frac{\Lambda_a}{\varepsilon+P}-\frac{\tau_{\pi\pi}}{4\tau_\pi}\frac{\Lambda_d}{\varepsilon+P}\geq0,\, (a\neq d)
    \label{eq:causal_n4}
\end{align}
\begin{align}
    n_5 &\equiv c^2_s + \frac{4}{3}\frac{1}{C_\eta} + \frac{1}{C_\zeta} + \left(\frac{2}{3}\frac{\lambda_{\pi\Pi}}{\tau_\pi} + \frac{\delta_{\Pi\Pi}}{\tau_\Pi} + c^2_s \right)\frac{\Pi}{\varepsilon+P} \nonumber \\
    & + \left( \frac{3\delta_{\pi\pi} + \tau_{\pi\pi}}{3\tau_\pi} + \frac{\lambda_{\Pi\pi}}{\tau_\Pi} + c^2_s\right) \frac{\Lambda_1}{\varepsilon + P} \geq 0,
    \label{eq:causal_n5}
\end{align}
\begin{align}
    n_6 &\equiv 1 - \left(c^2_s + \frac{4}{3}\frac{1}{C_\eta} + \frac{1}{C_\zeta}\right) \nonumber \\
    & + \left(1 - \frac{2}{3}\frac{\lambda_{\pi\Pi}}{\tau_\pi} - \frac{\delta_{\Pi\Pi}}{\tau_\Pi} - c^2_s\right)\frac{\Pi}{\varepsilon+P} \nonumber \\
    & + \left(1 - \frac{3\delta_{\pi\pi} + \tau_{\pi\pi}}{3\tau_\pi} - \frac{\lambda_{\Pi\pi}}{\tau_\Pi} - c^2_s\right)\frac{\Lambda_3}{\varepsilon + P} \geq 0.
    \label{eq:causal_n6}
\end{align}
In these conditions, we normalize the shear viscous tensor and bulk viscous pressure by the fluid cell's local enthalpy $\varepsilon + P$ to be unitless. The coefficients $C_\eta = \tau_\pi (\varepsilon + P)/\eta$ and $C_\zeta = \tau_\Pi (\varepsilon + P)/\zeta$ are the unitless ratios of shear and bulk viscosity to their relaxation times. Here $\Lambda_i$ are the eigenvalues of the shear stress tensor $\pi^{\mu\nu}$ with the ordering $\Lambda_3 > \Lambda_2 > \Lambda_1$. Because the shear stress tensor is transverse to the fluid velocity, $u_\mu \pi^{\mu\nu} = 0$, $\Lambda_0 = 0$. The rest three eigenvalues $\{\Lambda_i\} (i = 1, 2, 3)$ satisfy $\Lambda_1 + \Lambda_2 + \Lambda_3 = 0$ because of the shear stress tensor is traceless.

When any causality measure $n_i$ is violated, we propose to numerically reduce the size of $\Pi$ and $\pi^{\mu\nu}$ to recover causality. Since the causality measures $\{n_i\}$ are local and their values vary from cell to cell, we must impose the numerical regulation locally on individual fluid cells during the hydrodynamic simulations. In practice, we introduce a set of regulation factors $\alpha_i (i \in [1, 6])$ with $\alpha_i > 0$ for each necessary causality condition. We obtain $\alpha_i$'s value by solving $n_i(\alpha_i \Pi, \alpha_i \pi^{\mu\nu}) = 0$ for each of the necessary causality condition in Eqs.~\eqref{eq:causal_n1}-\eqref{eq:causal_n6}.
Once we get $\{\alpha_i\}$, we find $\alpha_{\rm min}= {\rm min}(\{\alpha_i\})$. If $\alpha_{\rm min} < 1$, we reduce the viscous variables by the factor of $\alpha_{\rm min}$:
\begin{equation}
    \bar{\Pi} = \alpha_{\rm min}\Pi,
    \label{eq:A7}
\end{equation}
and
\begin{equation}
    \bar{\pi}^{\mu\nu} = \alpha_{\rm min} \pi^{\mu\nu}.
    \label{eq:A8}
\end{equation}
The $\alpha_{\rm min}$ is chosen to satisfy all conditions after the numerical regulation. If $\alpha_{\rm min} \ge 1$ for some fluid cells, these cells do not violate the necessary causality conditions, whereas, with $\alpha_{\rm min} \sim 0$, the system is too far away from equilibrium to be described by the causal relativistic viscous hydrodynamics.

The standard numerical regularization schemes implemented in MUSIC can be cast in a similar fashion as in Eqs.~\eqref{eq:A7} and \eqref{eq:A8}~\cite{Schenke:2010rr, Shen:2014vra, Denicol:2018wdp, Chiu:2021muk}. In every fluid cell, we have the scale factors
\begin{align}
    \alpha_{\rm min, shear} = \mathrm{min}\left\{1, f_s(e)\sqrt{\frac{e^2 + 3P^2}{\pi^{\mu\nu} \pi_{\mu\nu}}} \right\}
\end{align}
for shear stress tensors and
\begin{align}
    \alpha_{\rm min, bulk} = \mathrm{min}\left\{1, f_s(e) \sqrt{\frac{e^2 + 3P^2}{3\Pi^2}} \right\}
\end{align}
for the bulk viscous pressure. In the standard numerical regulation schemes, the sizes of the regulation factors $\alpha_{\rm min}$ are controlled by an energy-density-dependent factor $f_s(e)$, 
\begin{align}
    f_s(e) = \chi_0 \left[\frac{1}{\exp\left(-\frac{e - e_0}{\xi_0}\right) + 1} -  \frac{1}{\exp\left(\frac{e_0}{\xi_0}\right) + 1} \right].
\end{align}
Here, the numerical parameters $\chi_0 = 1$, $e_0 = 0.02$~GeV/fm$^3$, and $\xi_0 = 0.05$~GeV/fm$^3$ were chosen to stabilize the event-by-event simulations. The energy dependence in $f_s(e)$ ensures that only strong regulations are imposed outside the physical freeze-out surfaces.

Our proposed numerical regulation procedure (in Eqs.~\eqref{eq:A7} and \eqref{eq:A8}) can effectively stabilize event-by-event hydrodynamic simulations in practice and impose the minimum required corrections to the bulk and shear viscous stress tensors of individual fluid cells to ensure causality.

\bibliography{Causality_Conditions, non-inspires}

\end{document}